\documentclass{XrU2005}
\usepackage{graphicx}

\title{XMM-Newton observations of the microquasars GRO J1655-40 and GRS 1915+105}
\author[1]{G. Sala}
\author[1]{J. Greiner}
\author[1]{F. Haberl}
\author[2]{E. Kendziorra}
\author[1]{K. Dennerl}
\author[1]{M. Freyberg}
\author[1]{G. Hasinger}
\affil[1]{Max-Planck-Institut f\"ur extraterrestrische Physik, Postfach 1312, D-85741 Garching, Germany}
\affil[2]{Institut f\"ur Astronomie und Astrophysik T\"ubingen, Sand 1, D-72076 T\"ubingen, Germany}

\begin{document}


\maketitle

\begin{abstract}
\vspace{-0.35cm}
We present results of a sequence of XMM-Newton observations of the two microqasars GRO J1655-40 and
GRS 1915+105. The observations were preformed using the EPCI pn camera in the Burst mode.
The observations of GRO J1655-40 in a bright state have made possible a substantial improvement in 
the calibration of the Burst mode, with determination of the rate dependence of the 
Charge Transfer Efficiency (CTE). We detect He-like Fe K-shell absorption
features in the EPIC-pn spectrum of GRO J1655-40, indicating the presence of a highly ionized absorber, 
and clear absorption features at 0.71 and 0.72 keV in the RGS spectrum, most probably identified as blueshifted 
Fe XVIII. 
\end{abstract}

\section{Introduction}

\vspace{-0.35cm}

Microquasars are accreting binary systems in our Galaxy ejecting jets at relativistic velocities. 
The microquasars GRO J1655-40 and GRS 1915+105 
were the two first superluminal sources discovered in our Galaxy. 
The dynamical mass of the central object, determined to be 7 M$_{\odot}$ for GRO J1655-40 \citep{oro97}
and 14 M$_{\odot}$ for GRS 1915+105 \citep{gre01}, indicates that it
is a black hole in both cases. GRO J1655-40 and GRS 1915+105 also share the peculiarity of being 
thought to contain a maximally spinning black hole \citep{zha97}.
ASCA observations of GRO J1655-40 in August 1994 and August 1995 provided the first 
detection of absorption lines in an accretion powered source \citep{ued98}.
The energy of the lines was found to depend on the X-ray intensity, 
being 6.95 keV (Fe XXVI K$\alpha$) at 2.2 Crab, and 6.63 and 7.66 keV
(Fe XXV K$\alpha$ and K$\beta$) at 0.27-0.57 Crab,   
revealing the presence of a highly ionized absorber.
Similar absorption features were also detected for GRS 1915+105 \citep{kot00}.

After 7 years of quiescence, GRO J1655-40 started a period of activity in February 2005, 
with RXTE/ASM showing a first outburst between March 10 and April 1, 
reaching $\sim$2 Crab, followed by a month and a half of increasing X-ray flux 
and a strong outburst on May 20, when the source reached more than 4 Crab. 
Here we present the results of four XMM-Newton observations of 
GRO J1655-40 performed on 27 February (40 ks, TOO), 
and on 14, 15 and 16 March 2005 (GT, 15 ks each).

\section{EPIC-pn CTE correction}

\vspace{-0.35cm}

\begin{figure}[!b]
\centering
\vspace{-0.4cm}
\includegraphics[width=1.1\linewidth]{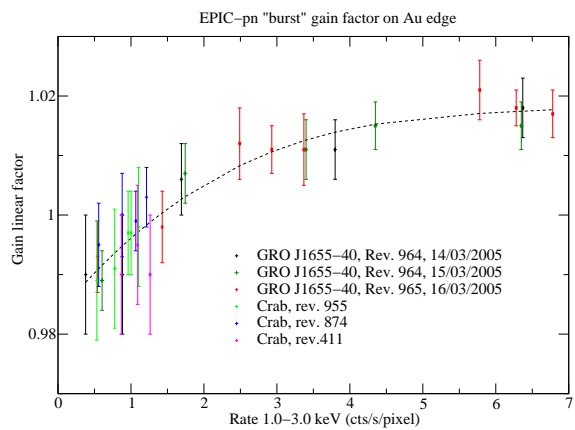}
\vspace{-0.8cm}
\caption{
Calibration of the gain linear factor \it{f} as a function 
of rate per pixel \it{r}.
\label{fig:fig1}}
\end{figure}

\begin{figure}[th]
\centering
\includegraphics[width=0.95\linewidth]{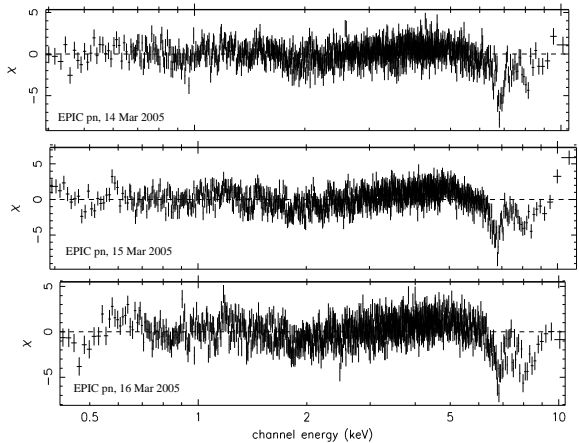}
\caption{EPIC pn residuals of GRO J1655-40 spectra of our three March 2005 obervations, 
after fitting an absorbed multi-temperature disk model.
\label{fig:fig2}}
\end{figure}

The bright state of GRO J1655-40 at the time of our XMM-Newton observations (almost twice
brighter than the Crab, the source used for Burst mode calibration)
has shown that the Charge Transfer Efficiency (CTE) in the Burst mode
has a stronger rate-dependence than previously modeled. An inaccurate 
calibration of the EPIC-pn CTE leads to a bad energy determination, which becomes
evident in the large residuals around the instrumental Si and Au edges.
From our GRO J1655-40 observations, the offset in energy has been found to be rate dependent, 
being stronger at the center of the PSF, 
which implies that it can not be directly corrected in the extracted spectrum.
We have determined the CTE gain
for different rates, selecting and evaluating the energy gain linear factor for 
spectra extracted from different regions of the detector  (Fig.~\ref{fig:fig1}). 
We have also included the Crab calibration observations to improve 
the determination of the dependence of the gain linear factor \textit{f} with 
the rate per pixel \textit{r} (cts/s/pixel), which we find can be approximated by 
$f=0.98+0.015r-2.2\times10^{-3}r^{2}+1.1\times10^{-4}r^{3}$.
After correcting the event tables with this linear gain, no more 
residuals appear around the Au and Si edges.

\section {GRO J1655-40 and GRS 1915+105}

\vspace{-0.35cm}

\begin{figure}[ht]
\centering
\includegraphics[width=0.95\linewidth]{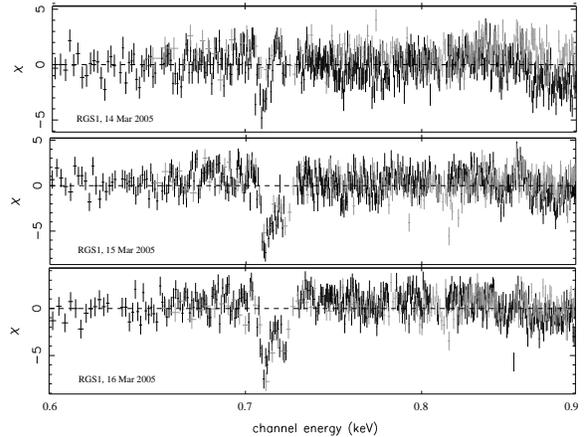}
\caption{RGS 1 order 1 (black) and 2 (grey) residuals for our March 2005 observations of GRO J1655-40.
\label{fig:fig3}}
\end{figure}

For our first XMM-Newton observation of GRO J1655-40, performed on 
27 February 2005, some days before the start of the first outburst, 
a simultaneous RXTE observation is available, which has allowed to determine
the X-ray spectrum up to 60 keV. The simultaneous fit to XMM-Newton and RXTE/PCA data
shows that the spectrum can be modeled by a multi-temperature accretion disk, 
plus a power law with index $\sim$1.5.
During our March observations (with exposures of $\sim$15ks), 
performed close to the maximum of the first outburst,
the spectra of the EPIC pn camera, 
which was used in the Burst mode, show that the 
emission is dominated by the multi-temperature accretion disk 
component, with kT$_{in}$=1.2-1.3 keV, hotter than the 
typical temperature observed during the 1996-1997 outburst \citep{sob99}.
We detect two absorption features at 6.8 and 8.0 keV, 
corresponding to Fe XXV K$\alpha$ and K$\beta$ 
absorption lines (Fig.~\ref{fig:fig2}). 
The simultanous RGS data provide the first high resolution spectra of GRO J1655-40,  
showing clear absorption lines at 0.71 and 0.72 keV (Fig.~\ref{fig:fig3}),
which could be identified either as OVII at zero velocity, or as 
a blueshifted Fe XVIII L-shell doublet.  The blueshift would indicate in 
this case an outflowing absorber at 3000 km/s. 

\begin{figure}[ht]
\smallskip
\centering{
\includegraphics[width=0.83\linewidth]{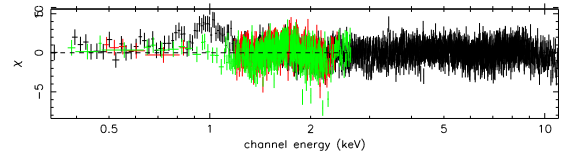}}
\caption{EPIC pn and RGS residuals of the GRS 1915+105 spectrum.}
\label{fig:fig4}
\end{figure}

The absence of other absorption features in the RGS spectrum is 
puzzling:
if the lines at $\sim$0.7 keV correspond to OVII K-shell at zero velovity, 
we would expect to see an even stronger absorption 
of OVII K$\alpha$ at $\sim$0.57 keV that is not present; 
and if they are Fe XVIII absorption, Fe XVII should be also present.
In addition, the lack of observable OVIII and the presence of Fe XXV 
can only be explained by a highly ionized gas,
with temperatures higher than $kT\sim1.7$ keV. But at higher temperatures,  
Fe XVIII would not be expected to be present, constraining the possible
temperature of the absorber to a narrow range, higher than the disk temperature
derived from its thermal emission ($kT\sim1.2-1.3$ keV).

XMM-Newton observed GRS 1915+105 on 3 May 2004, with 20 ks exposure time, and the EPIC-pn camera in the 
Burst mode. The EPIC and RGS spectra can be fitted with an absorbed power law, and no evident 
absorption features are observed. Since the source was not so bright as GRO J1655-40, 
no CTE problems are found and no residuals appear around the Si and Au edges.
Nevertheless, the best-fit leaves an excess in the EPIC-pn spectrum around 1 keV
not observed in the two RGS spectra, which could indicate some calibration problem.

{\bf ACKNOWLEDGEMENTS} We thank J. Vink, J. Kaastra and Y. Ueda for useful discussions in El Escorial.
GS acknowledges a MEC postdoctoral fellowship.

\end{document}